\documentstyle[aps,epsf]{revtex}
\begin{document}

\title{Winding angle distribution of 2D random walks with traps}

\author{K. Samokhin$^*$}

\address{Cavendish Laboratory, University of Cambridge, Madingley Road,
         Cambridge CB3 0HE, UK}
  
\date{\today}

\maketitle

\begin{abstract}
We study analytically the asymptotic behaviour of the average probability 
${\cal P}(n,t)$ for the trajectory of a 2D Brownian particle wandering 
in the presence of randomly distributed traps to wind $n$ times around a
given point after a time $t$. It is shown that ${\cal P}(n,t)\sim 
\exp(-c\sqrt{t})(1+x^2)^{-1}$ with $x\sim n/\sqrt{t}$, where the first 
exponent represents a well-known long-time tail of the probability that a 
particle will not be trapped.
\end{abstract}
\bigskip\bigskip

The properties of random walks with various topological constraints has 
attracted a great deal of theoretical interest for many years. Apart from 
apparent practical relevance to the physics of polymers or Abrikosov vortex 
lines in superconductors (see, e.g., Ref. \cite{Wiegel86}), studies 
of such random walks make profound and intimate connections to many 
beautiful mathematical results. Perhaps, the most prominent example is the 
problem of the winding angle distribution in two dimensions. The winding angle
of a planar random walk is, by definition, the total continuous angle   
$\theta(t)=2\pi n(t)$ swept by a Brownian particle around a prescribed point 
after a time $t$. It was found by Spitzer \cite{Spit58} that the asymptotic
probability to wind $n$ times is given by a Cauchy law:
\begin{equation}
\label{Spitzer}
 {\cal P}(n,t)\sim\frac{1}{1+x^2},\quad x\sim\frac{n}{\ln t},\qquad
 \mbox{at }t\to\infty.
\end{equation}
This result was later confirmed by many authors by employing different 
techniques (see, e.g., Refs. \cite{Edw67,CDM93,DK96}). 
In general, one could also ask how many times a particle has wound around a 
set of $N$ prescribed points \cite{PY86} or an excluded disk in a 2D plane 
\cite{RH87}.

Let us now suppose that our Brownian particle can not wander freely, but 
instead can be irreversibly trapped by the impurities located at some
randomly distributed points in a plane. It is known that the properties
of such a system differ drastically from those of an ideal random walk. 
For instance, the survival probability is given by $P(t)\sim\exp(-c\sqrt{t})$
\cite{BV74}, while the mean square displacement is sub-diffusional:
$\langle r^2\rangle\sim\sqrt{t}$ \cite{GP82}. The purpose of the present 
article is to discover how the presence of traps affects the winding angle 
distribution function.

The probability distribution for a random walk starting at a point 
${\bf r}'$ to end at a point ${\bf r}$ after a time $t$ satisfies the 
diffusion equation
\begin{equation}
\label{diff_eq}
 \frac{\partial P}{\partial t}=D\nabla^2P-U({\bf r})P.
\end{equation}
Here $D$ is the diffusion coefficient and $U({\bf r})=U_0\sum_i\delta({\bf r}-
{\bf R}_i)$ is the random ``potential'', which is the probability per unit 
time for a particle to be trapped ($U_0>0$). The positions ${\bf R}_i$ of 
point-like traps are distributed uniformly in a plane according to the 
Poisson law with mean density $\rho$. The solution of Eq. (\ref{diff_eq})
is given by the Wiener path integral formula:
\begin{equation}
\label{Wiener}
 P({\bf r},t;{\bf r}',0|U)=\int_{{\bf r}(0)={\bf r}'}^{{\bf r}(t)={\bf r}} 
   {\cal D}{\bf r}(\tau)\;\exp\left\{-\int_0^t d\tau\;\left(\frac{1}{4D} 
    \dot{\bf r}^2(\tau)+U({\bf r}(\tau))\right)\right\}.
\end{equation}
The probability for a closed random walk with ${\bf r}={\bf r}'$ of ``length''
$t$ to wind $n$ times around the origin in a given distribution of traps 
can be calculated by inserting a $\delta$-function constraint \cite{Edw67} 
in (\ref{Wiener}), so that
\begin{equation}
 {\cal P}(n,t|U)=\left\langle \delta\left(n-\frac{1}{2\pi}\int_0^td\tau\;
  \dot\theta(\tau)\right)\right\rangle_{P({\bf r},t;{\bf r},0|U)},
\end{equation}
where $\theta(t)$ is the angle between the radius-vector ${\bf r}(t)$ and 
some fixed direction in the plane (note that $n$ can be non-integer). 
Writing the $\delta$-function as an integral over an auxiliary variable $p$, 
we arrive at
\begin{equation}
 {\cal P}(n,t|U)=\int_{-\infty}^\infty dp\;e^{2\pi {\rm i}pn}
  \int_{{\bf r}(0)={\bf r}}^{{\bf r}(t)={\bf r}}{\cal D}{\bf r}(\tau)\;
   \exp\left\{-\int_0^t d\tau\;\left(\frac{1}{4D} \dot{\bf r}^2(\tau)
   +U({\bf r}(\tau))+{\rm i}p\dot\theta(\tau)\right)\right\}. 
\end{equation}
If one assumes that the starting point ${\bf r}$ is not fixed, then the path 
integral on the right-hand side is nothing but the partition function at 
inverse ``temperature'' $1/T=t$ of a particle of unit charge and mass 
$m=(2D)^{-1}$ moving in a random potential and in a solenoid field localised 
at the origin, the solenoid carrying a flux $\phi=-2\pi p$. For the average 
probability we then have
\begin{equation}
\label{gen_P}
 {\cal P}(n,t)\equiv\langle {\cal P}(n,t|U)\rangle_U=\int_{-\infty}^\infty
 \frac{d\phi}{2\pi}\int_0^\infty dE\;e^{-{\rm i}\phi n}e^{-Et}N(E,\phi),
\end{equation}
where $N(E,\phi)$ is the average density of states. Since we are interested
in the asymptotic behaviour of ${\cal P}(n,t)$ at large $t$, all we have to do 
is to calculate the asymptotics of $N(E,\phi)$ at small $E$, which is called 
``Lifshitz tail'' \cite{Lif65}. 

From the analysis of a random walk with traps but without solenoid it is 
known that at $E\to 0$ the main contribution to the density of states comes 
from the large regions in real space which are free of traps. The probability 
to find such a region of area $S$ is exponentially small: 
$p(S)\sim e^{-\rho S}$. From the elementary quantum mechanics we know that 
the ground state energy of a particle in a 2D potential well with radius $R$ 
is given by $E(R)\sim DR^{-2}\sim DS^{-1}$. Therefore, $S(E)\sim D/E$, and the 
density of states is $N(E)\sim p(S(E))\sim \exp(-{\rm const}\;\rho D/E)$. 
More formally, such exponentially small tails of the density of states 
correspond to the contribution of instantons \cite{Lang67}, which are 
spatially localised solutions of the saddle-point equations in the 
functional-integral representation of the problem \cite{FL75,Lub84}. 
Our strategy is as follows. First, we formulate the problem in the language 
of quantum field theory with some effective action $S$. 
Then, we find an explicit form of the instanton solution. The last step is to 
calculate the asymptotic behaviour of ${\cal P}(n,t)$ at large $t$ due to the 
instanton contributions.  
 
The Schr\"odinger equation for a quantum particle moving in the field of a 
solenoid and in the potential $U({\bf r})$ is as follows:
\begin{equation}
 H\psi\equiv D(-{\rm i}\nabla-{\bf A}({\bf r}))^2\psi+U({\bf r})\psi=E\psi,
\end{equation}
where $A_\theta=\phi/2\pi r$ is the vector potential created by the solenoid.
The density of states is proportional to the imaginary part of the 
Green function $G_E({\bf r},{\bf r})=\langle{\bf r}|(E-H+{\rm i}0)^{-1}|{\bf r}
\rangle$, which can be calculated by standard means of the quantum field 
theory. Using the replica trick, the disorder average of the (Euclidean)
generating functional can be performed, and we have in the limit $n\to 0$
\begin{eqnarray}
\label{G_R}
 G_E({\bf r},{\bf r})=\int{\cal D}^2\mbox{\boldmath $\varphi$}({\bf r})\;
  e^{-S[\mbox{\boldmath $\varphi$}({\bf r})]} \varphi_1({\bf r})
  \bar\varphi_1({\bf r}),
\end{eqnarray}
where $\mbox{\boldmath $\varphi$}$ is an $n$-component Bose field and 
${\cal D}^2\mbox{\boldmath $\varphi$}=\prod_{a=1}^n{\cal D}\bar\varphi_a
{\cal D}\varphi_a$. The action is
\begin{equation}
\label{action}
 S=\int d^2r\; \Bigl\{ \bar{\mbox{\boldmath $\varphi$}}\Bigl(-E+D(-{\rm i}
 \nabla-{\bf A})^2\Bigr)\mbox{\boldmath $\varphi$}+\rho\Bigl(1-e^{-U_0
  \bar{\mbox{\boldmath $\varphi$}}\mbox{\boldmath $\varphi$}}\Bigr)\Bigr\}.
\end{equation}
At $E<0$ the action is always positive, so that the field theory is stable 
and the imaginary part is zero. At $0<E\ll E_{\rm c}=\rho U_0$ ($E_{\rm c}$ 
being the mean value of the random potential) there is a metastable vacuum 
state $\mbox{\boldmath $\varphi$}=0$. In this case, the small-$E$ asymptotics 
of the imaginary part of the Green function is determined by a non-trivial 
saddle point of the action \cite{Lang67} and, with exponential accuracy,
\begin{equation}
\label{DoS_inst}
 N(E,\phi)\sim e^{-S_{\rm inst}(E,\phi)}.
\end{equation}
Due to the rotational symmetry of Eq. (\ref{action}) in the $n$-dimensional 
replica space, the saddle-point solution (instanton) has the form
\begin{equation}
\label{sp_sol}
 \varphi_a({\bf r})=\varphi({\bf r})e_a,
\end{equation} 
where $e_a$ is the $a$-th component of an arbitrary $n$-component unit vector.
From (\ref{action}) and (\ref{sp_sol}) we obtain the following equation for 
the function $\varphi({\bf r})$ which is assumed to be rotationally invariant 
in real space (i.e. $\varphi({\bf r})=\varphi(r)e^{{\rm i}m\theta}$ with 
$m=0$):
\begin{equation}
\label{inst_eq}
 -D\frac{1}{r}\frac{d}{dr}\left(r\frac{d}{dr}\right)\varphi+
  D\frac{\nu^2}{r^2}\varphi+E_{\rm c}e^{-U_0\varphi^2}\varphi=E\varphi,
\end{equation}
where $\nu=|\phi|/2\pi$. Let us introduce the dimensionless variables:
$$
 r=\xi x,\quad \varphi(x)=U_0^{-1/2}f(x).
$$
Here $\xi^2=D/E$ is a characteristic scale of the problem, which is 
nothing but the typical length of diffusion in time $t=E^{-1}$. 
Eq. (\ref{inst_eq}) can then be written as
\begin{equation}
\label{gen_eq}
 -\frac{1}{x}\frac{d}{dx}\left(x\frac{d}{dx}\right)f+\frac{\nu^2}{x^2}f+
  \alpha^2e^{-f^2}f=f,
\end{equation}
where $\alpha^2=E_{\rm c}/E$. 

Since Eq. (\ref{gen_eq}) is a non-linear differential equation, we are able
to find only an approximate solution.
As shown in Appendix, at $E\ll E_{\rm c}$ (i.e. $\alpha\gg 1$)
one can replace the ``potential'' $V(f)=\alpha^2e^{-f^2}$ in Eq. (\ref{gen_eq})
by the potential well with infinitely high walls, having the shape of a
coaxial ring with the inner and outer radii $x_1$ and $x_2$ respectively. 
Then the instanton solution inside the ring satisfies the Schr\"odinger 
equation for a particle in the solenoid field, so that
\begin{equation}
 f(x)=A_1J_\nu(x)+A_2Y_\nu(x),\quad \mbox{at }x_1<x<x_2,
\end{equation}
where $J_\nu(x)$ and $Y_\nu(x)$ are the Bessel functions of the first and 
second kind respectively. The positions of the matching points $x_1(\nu)$ 
and $x_2(\nu)$ are to be determined from the following equations: 
\begin{equation}
\label{fin_eqs}
 \left\{ \begin{array}{l}
 x_1F_\nu(x_1)J_\nu(x_1)=-x_2J_\nu(x_2), \\
 x_1F_\nu(x_1)Y_\nu(x_1)=-x_2Y_\nu(x_2),
 \end{array} \right.
\end{equation}
where 
$$
  F_\nu(x_1)=\frac{I'_\nu(\alpha x_1)}{I_\nu(\alpha x_1)}=
  \frac{\nu}{\alpha x_1}+\frac{I_{\nu+1}(\alpha x_1)}{I_\nu(\alpha x_1)}
$$
(the details of derivation can be found in Appendix). These equations are 
valid for $E\ll E_{\rm c}$ and arbitrary $\nu$. Going back to the dimensional 
variables, it is easy to convince oneself that the instanton action coincides 
with the area of the ring:
\begin{equation}
\label{S_inst}
 S_{\rm inst}=2\pi\rho\int_{r_1}^{r_2}dr\;r=\pi\rho\xi^2
       (x_2^2(\nu)-x_1^2(\nu)).
\end{equation}
In the absence of solenoid (i.e. at $\nu=0$), the solution of Eqs. 
(\ref{fin_eqs}) is $x_1=0$, $x_2=a$, where $a\approx 2.405$ is the first zero 
of the function $J_0(x)$. After substitution in (\ref{S_inst}), the
Lifshitz result \cite{Lif65} is recovered. 

At $\nu\neq 0$, due to fast oscillations of the phase factor $e^{-{\rm i}\phi 
n}$ in (\ref{gen_P}), the main contribution to the integral comes from small 
$\phi$, which allows one to use a perturbative expansion in powers of 
$\nu$. We seek a solution of (\ref{fin_eqs}) in the form 
\begin{equation}
\label{expansion}
 x_1=\delta x_1(\nu),\quad x_2=a+\delta x_2(\nu)\quad (\alpha\delta x_1\to 0). 
\end{equation}
The Bessel functions can be expanded in powers of their index \cite{AS65}:
$$
 J_\nu(x)=J_0(x)+\frac{\pi\nu}{2}Y_0(x)+O(\nu^2),\qquad
 Y_\nu(x)=Y_0(x)-\frac{\pi\nu}{2}J_0(x)+O(\nu^2).
$$
However, one should be careful in dealing with such expressions, 
since the Bessel functions are not analytical at $x=0$, so that we are able 
to safely expand only the right-hand sides of Eqs. (\ref{fin_eqs}). 
Using the small-$x$ expansions of the Bessel functions on the left-hand sides,
we obtain, in the leading order in $\delta x_{1,2}$ and $\nu$:
$$
 \left\{ \begin{array}{l}
  \displaystyle \frac{1}{\alpha}\frac{1}{\Gamma(\nu)}\left(
   \frac{\alpha\delta x_1}{2}\right)^\nu=aJ_1(a)\delta x_2-\frac{\pi a}{2}
   Y_0(a)\nu, \\
  \displaystyle \frac{1}{\alpha}\frac{\Gamma(\nu+1)}{\pi}\left(
  \frac{\alpha\delta x_1}{2}\right)^{-\nu}=aY_0(a),
 \end{array} \right.
$$
where $\Gamma(x)$ is the Gamma function. Therefore, the solution of 
Eqs. (\ref{fin_eqs}) looks as follows:
\begin{equation} 
 \left\{ \begin{array}{l}
  \displaystyle x_1=\frac{2}{\alpha}\left(\frac{\pi aY_0(a)\alpha}{\Gamma
  (\nu+1)}\right)^{-1/\nu}\simeq 
  \frac{2}{\alpha}(\pi aY_0(a)\alpha)^{-1/\nu}, \\
  \displaystyle x_2=a+\left(\frac{\pi Y_0(a)}{2J_1(a)}+\frac{1}{\pi a^2J_1(a)
  Y_0(a)\alpha^2}\right)\nu\simeq a+\frac{\pi Y_0(a)}{2J_1(a)}\nu.
 \end{array} \right.
\end{equation} 
At $\nu\to 0$ $x_1$ vanishes faster than $\delta x_2$, so that its contribution
to the instanton action can be neglected. From (\ref{S_inst}), we then obtain
at $\phi\to 0$:
\begin{equation}
\label{result_action} 
 S_{\rm inst}(E,\phi)=\frac{\pi\rho Da^2}{E}\left(1+b|\phi|+O(\phi^2)\right),
\end{equation}
where $b=\frac{Y_0(a)}{2aJ_1(a)}\approx 0.204$. A non-analytical dependence 
on the magnetic flux is related to the fact that one can not regard the 
solenoid field as a small perturbation due to $r^{-2}$-singularity at small 
$r$. 

Finally, we see from (\ref{DoS_inst}) that the asymtotics of the average 
density of states in the presence of solenoid is given by
\begin{equation}
\label{DOS}
 N(E,\phi)\sim\exp\left(-\frac{\pi\rho Da^2(1+b|\phi|)}{E}\right).
\end{equation}
To calculate the pre-exponential factor, one should make an expansion around 
the instanton configuration and integrate over all non-zero modes. We, 
however, shall not proceed in this way further and restrict ourselves by the 
exponential accuracy. After substitution of (\ref{DOS}) in (\ref{gen_P}), 
we arrive at
$$
 {\cal P}(n,t)\sim\int_{-\infty}^\infty d\phi\int_0^\infty dE\;e^{-{\rm i}
  \phi n}e^{-Et}\exp\left(-\frac{\pi\rho Da^2(1+b|\phi|)}{E}\right). 
$$
At large $t$ the integral over $E$ can be calculated by the method of 
steepest descent, resulting in
\begin{eqnarray}
\label{result}
 {\cal P}(n,t)&\sim& \int_{-\infty}^\infty d\phi\;e^{-{\rm i}\phi n}
  \exp\Bigl(-2a\sqrt{\pi\rho Dt}\sqrt{1+b|\phi|}\Bigr) \nonumber \\
  &\sim& \exp(-2a\sqrt{\pi\rho Dt})\left(1+\frac{n^2}{c\rho Dt}
   \right)^{-1},
\end{eqnarray}
where $c=\pi a^2b^2\approx 0.756$. In calculating the last integral we 
used the fact that the main contribution comes from small values of 
the flux, which justifies an expansion in powers of $\phi$.  
The exponential factor on the right-hand side of (\ref{result})
represents the asymptotic probability for a particle without solenoid
to survive after a time $t$ and coincides with the result of Balagurov and
Vaks \cite{BV74}. The second factor can thus be interpreted as the
conditional probability for a particle which has survived
to wind $n$ times around the origin. 

It is expedient to compare our results with what is known for other similar 
systems. For an ideal random walk without traps, the scaling variable is 
$x=n/\ln t$, whose asymptotic distribution is given by Spitzer's law 
(\ref{Spitzer}). For a self-avoiding random walk without traps, the scaling 
variable $x=n/\sqrt{\ln t}$ has a Gaussian distribution \cite{FPR84,DS88}. 
In that case, due to the hard-core repulsion, the trajectory wanders farther 
away from the origin than does an ideal random walk, which reduces the 
winding number. In our case, we see that the scaling variable is 
$x=n/\sqrt{t}$, and the asymptotic distribution obeys a Cauchy law. 
The increase of the winding number can be qualitatively understood as follows.
We are considering the conditional probability, which implies that the 
particle has survived until the time $t$. This, in turn, means that it spent 
much of its life in a finite region of the plane almost free of traps and 
thus has never wandered too far away from the starting point. Such a 
restriction obviously results in increasing entanglement. \\    

The author would like to thank D. E. Khmel'nitskii for numerous stimulating
discussions. The work was financially supported by the EPSRC grant no 
RG 22473. 

\appendix
\section*{}

\begin{figure}
\begin{center}
\leavevmode
\epsfxsize=0.7 \linewidth
\epsfbox{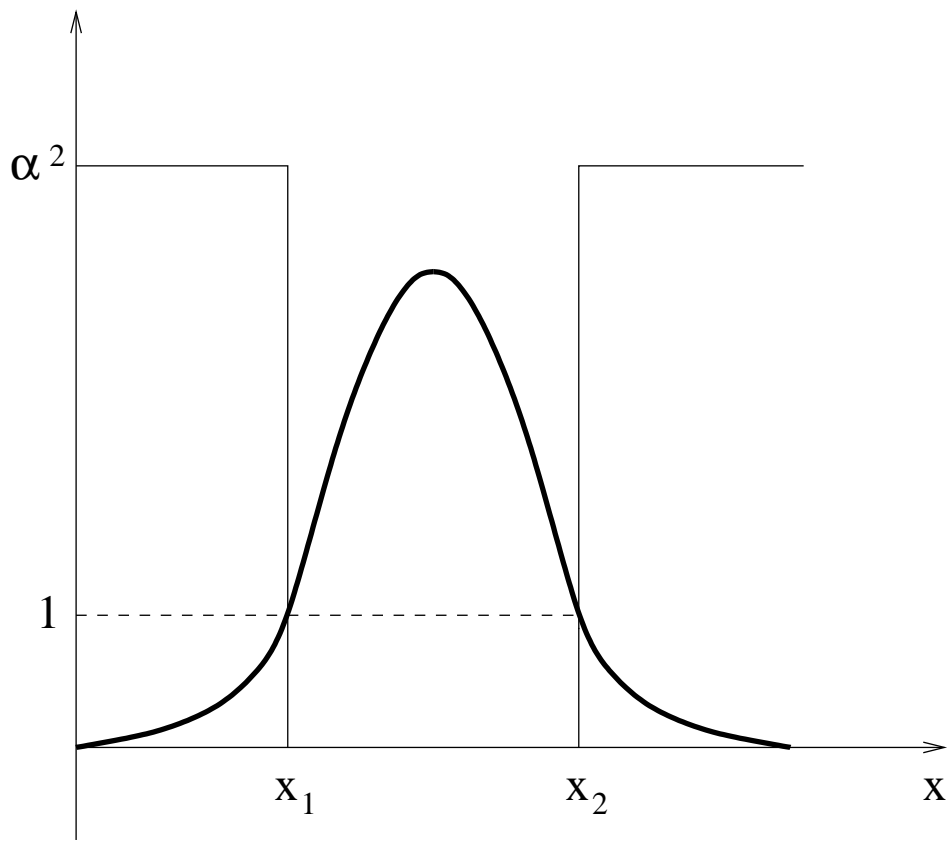}
\caption{The effective potential $V_{\rm eff}(f(x))$ (light line) and the 
         instanton solution $f(x)$ (heavy curve) as functions of 
         $x=r/\xi$ ($\alpha^2=E_{\rm c}/E\gg 1$).}
\label{instanton_fig}
\end{center}
\end{figure}

We replace the ``potential'' $V(f)=\alpha^2e^{-f^2}$ in the nonlinear equation
(\ref{gen_eq}) by a piecewise constant effective potential:
\begin{equation}
 V_{\rm eff}(f)=\left\{ \begin{array}{ll}
  \alpha^2\ &, \mbox{ at }f<1, \\
  0\ &, \mbox{ at }f>1.
  \end{array} \right.  
\end{equation}
Then the solution is a piecewise continuous function:
\begin{equation}
 f(x)=\left\{ \begin{array}{ll}
  f_1(x)\ ,& \mbox{ at }0<x<x_1, \\
  f_2(x)\ ,& \mbox{ at }x_1<x<x_2, \\
  f_3(x)\ ,& \mbox{ at }x_2<x,
 \end{array} \right.
\end{equation}
where the functions $f_i(x)$ obey the following linear equations:
\begin{equation}
\label{eqs}
\left\{ \begin{array}{l}
 \displaystyle -\frac{1}{x}\frac{d}{dx}\left(x\frac{d}{dx}\right)
 f_{1,3}+\frac{\nu^2}{x^2}f_{1,3}+\alpha^2f_{1,3}=f_{1,3}, \\
 \displaystyle -\frac{1}{x}\frac{d}{dx}\left(x\frac{d}{dx}\right)f_2+
  \frac{\nu^2}{x^2}f_2=f_2.
\end{array} \right.
\end{equation}
The solution $f(x)$ and its derivatives must be continuous functions of $x$ at
$x=x_{1,2}$. The positions of the matching points $x_1$ and $x_2$ are 
determined from the conditions $f_1(x_1)=f_2(x_1)=1$ and 
$f_2(x_2)=f_3(x_2)=1$ (see Fig. \ref{instanton_fig}).

The solution of Eqs. (\ref{eqs}) looks as follows:
\begin{equation}
\label{gen_solutions}
\left\{ \begin{array}{l}
 f_1=C_1I_\nu(\alpha x), \\
 f_2=C_2^{(1)}J_\nu(x)+C_2^{(2)}Y_\nu(x), \\
 f_3=C_3K_\nu(\alpha x),
\end{array} \right.
\end{equation}
where $I_\nu(x)$ and $K_\nu(x)$ are the Bessel functions of imaginary 
argument. The boundary conditions read
\begin{equation}
\label{bcs}
\left\{ \begin{array}{l}
 f_1(x_1)=f_2(x_1)=1,\ f_2(x_2)=f_3(x_2)=1, \\
 f'_1(x_1)=f'_2(x_1),\ f'_2(x_2)=f'_3(x_2).
\end{array} \right.
\end{equation}
Substituting (\ref{gen_solutions}) in (\ref{bcs}), we obtain a 
system of six transcendent equations to determine $C_1$, $C_2^{(1,2)}$, 
$C_3$, $x_1$ and $x_2$. After changing notations $C_2^{(1,2)}=\alpha A_{1,2}$,
the equations for $A_{1,2}$ and $x_{1,2}$ take the form
\begin{equation}
\label{A_eqs}
\left\{ \begin{array}{l}
  A_1J_\nu(x_1)+A_2Y_\nu(x_1)=\alpha^{-1}, \\
  A_1J_\nu(x_2)+A_2Y_\nu(x_2)=\alpha^{-1}, \\
  A_1J'_\nu(x_1)+A_2Y'_\nu(x_1)=
    \frac{I'_\nu(\alpha x_1)}{I_\nu(\alpha x_1)}, \\
  A_1J'_\nu(x_2)+A_2Y'_\nu(x_2)=
    \frac{K'_\nu(\alpha x_2)}{K_\nu(\alpha x_2)}.
\end{array} \right.
\end{equation}
In the limit $\alpha\gg 1$ the right-hand sides of the first two equations 
vanish. If one assumes that $x_2\sim 1$, then in the same limit the right-hand 
side of the last equation tends to $-1$. Excluding $A_{1,2}$ from 
(\ref{A_eqs}), we arrive at Eqs. (\ref{fin_eqs}).  

It is also worth explaining why we choose to split the plane into three 
different regions. If there were no solenoid ($\nu=0$), then $x_1=0$ and the 
instanton solution would be given by the Bessel function $J_0(x)$ which tends 
to a constant at $x\to 0$. However, if $\nu\neq 0$, then the naive assumption 
that $x_1=0$ and $f(x)\sim J_\nu(x)$ is not consistent with the condition 
that $f(x)>1$ everywhere inside the potential well, since $J_\nu(x)\sim x^\nu$
at $x\to 0$. For this reason one has to introduce the inner matching point 
$x_1\neq 0$.

\end{document}